\documentclass[aps,prl,twocolumn, superscriptaddress,letterpaper]{revtex4}
\usepackage{graphicx}
\usepackage{amssymb,amsmath}
\usepackage{float}
\usepackage{bm}
\usepackage{siunitx}
\usepackage{textcomp}

\usepackage{setspace}
\usepackage{parskip}
\usepackage{xcolor}
\definecolor{midnightblue}{cmyk}{1,1,0,0.1}
\definecolor{forestgreen}{cmyk}{0.76,0,0.26,0.5}

\usepackage{epsfig}

\usepackage{hyperref}
\hypersetup{
    bookmarks=true,         
    unicode=false,          
    pdftoolbar=true,        
    pdfmenubar=true,        
    pdffitwindow=false,     
    pdfstartview={FitH},    
    pdftitle={My title},    
    pdfauthor={Author},     
    pdfsubject={Subject},   
    pdfcreator={Creator},   
    pdfproducer={Producer}, 
    pdfkeywords={keyword1} {key2} {key3}, 
    pdfnewwindow=true,      
    colorlinks=true,       
    linkcolor=midnightblue,          
    citecolor=magenta,        
    filecolor=midnightblue,      
    urlcolor=midnightblue,          
}

\begin{document}

\title{Electronic Structure of Single-Twist Trilayer Graphene}

\author{Xiao Li}
\affiliation{NNU-SULI Thermal Energy Research Center  $\&$ Center for Quantum Transport and Thermal Energy Science, School of Physics and Technology, Nanjing Normal University, Nanjing 210023, China }
\affiliation{Department of Physics, University of Texas at Austin, Austin, Texas 78712, USA}

\author{Fengcheng Wu}
\affiliation{Condensed Matter Theory Center and Joint Quantum Institute, Department of Physics, University of Maryland, College Park, Maryland 20742, USA}

\author{Allan H. MacDonald}
\affiliation{Department of Physics, University of Texas at Austin, Austin, Texas 78712, USA}

\begin{abstract} 
Small-twist-angle bilayer graphene supports  
strongly correlated insulating states and superconductivity.  
Twisted few-layer graphene systems are likely to open up new 
directions for strong correlation physics in moir\'e superlattices.
We derive and study moir\'e band models that describe the electronic structure of 
graphene trilayers in which one of the three layers
is twisted by a small angle relative to perfect AAA, ABA, or ABC stacking 
arrangements.  We find  that the electronic structure depends very 
strongly on the starting stackings arrangement and on which layer is twisted.
We identify ABA stacking with a middle-layer twist as a promising system
for itinerant electron magnetism or even more robust superconductivity,
because it exhibits both large and small velocity bands at energies near the 
Fermi level. 
\end{abstract}

\maketitle
\textcolor{forestgreen}{\emph{\textsf{Introduction}.}}---A small relative twist between van der Waals layers  produces a  long-wavelength moir\'e pattern  \citep{Liu:2014, Cao:2016iea, Kim:2017gc, Li2010, Cao:2018kn,Cao:2018ff, Chen:2016, Zuo:2018jh}. Recently moir\'e superlattice systems have been realized experimentally in a number of two-dimensional layered materials, including graphene and transition metal dichalcogenides thanks to relatively weak interlayer interaction. 
Surprisingly, twisting layers to produce moir\'e patterns has turned out to be 
a powerful platform for creating and designing exotic electronic states, 
including ones with non-trivial band topology \citep{Tong:2016kha, Wu:2017du},
 unconventional superconductivity \citep{Cao:2018ff}, bands of 
 interlayer and intralayer excitons \citep{Yu:2015, Wu:2017du} and magnetization \citep{Alex:2018, Wu:2018ic}. 
For the specific case of twisted bilayer graphene (TBG), 
vanishing renormalized Dirac velocities and  flat electronic bands 
have been predicted at a series of small magic twist angles \citep{Bistritzer:2011ho}. 
Magic angle bilayer systems exhibit Mott insulating phases and superconductivity\citep{Cao:2018ff, Cao:2018kn},
and it is natural to expect modified but related behavior in all graphene multilayers when twists are present.
The electronic structure of  twisted trilayer graphene (TTG), for example, 
has been addressed in recent studies which focused on band topology, 
band evolution, and optical properties, 
among other quantities.\citep{Correa:2013dd, SuarezMorell:2013eta,Qiao:2014ez, Zuo:2018jh, Amorim:2018wb, Liu2019, Ma2019, Christophe2019}.  We focus here on a relatively simple case, twisted trilayer graphene (TTG), with 
a twist in  one layer only, which has not been investigated systematically.  
Single-twist TTG systems are distinguished by the starting crystalline stacking 
arrangement, and by which layer of the three layers is twisted.\citep{Aoki:2007,Bao:2017}.
The new features present in the trilayer system have generalizations to thicker 
multi-layer stacks.  

\begin{figure}[h!]
\centering
\includegraphics[width=7.6cm]{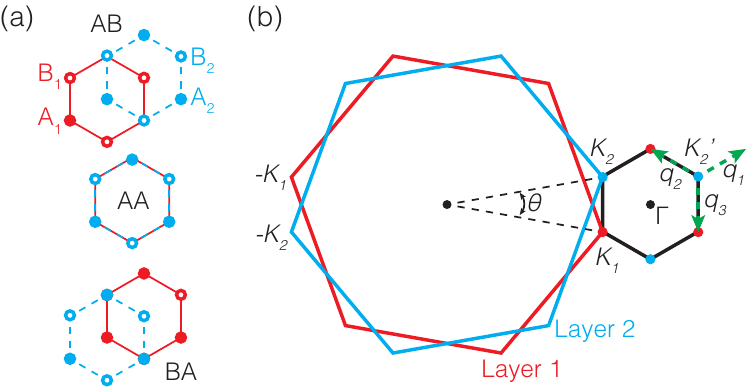}
\caption{ Real and reciprocal space of bilayer graphene. (a) Atomic structure of aligned bilayers with AB, AA and BA stacking.  Whereas the atoms of two layers are completely superimposed in AA stacking,  AB and BA stacking is characterized 
by having half of the atoms in one layer over the center of a hexagon in the other layer. 
(b) The moir{\'e}  Brillouin zone (Black) defined by  Brillouin zones of two graphene 
layers (Red and Blue). The momentum boosts, $\bm q_{1,2,3}$,  produced by 
tunneling between twisted layers are denoted by green arrows.}
\label{fig:1}
\end{figure}

In this paper, we study the electronic properties of TTGs devices
in which one of the three layers is twisted through a small angle relative to 
perfect  ABA,  ABC or AAA stacked structures, 
using a continuum moir{\'e} Hamiltonian. 
We demonstrate that different single-twist TTG structures
have quite different electronic structures. 
This property is in contrast to TBG, in which top-layer and bottom-layer twists relative 
to AB, AA or BA stacking [See Fig. \ref{fig:1}(a)] all lead to equivalent structures.  
More interestingly, ABA-stacked TTG with a  twist in the middle layer exhibit flat bands 
at twist angles that are larger than for TBG and, because of mirror symmetry with respect to the middle layer,
supports both large and small velocity bands near the Fermi level,   Both the number and the value of magic angles increase with the number of layers in twisted few-layer graphene. These  exotic characteristics identify twisted graphene layers with
mirror symmetry as a promising system in which to seek new strong correlation physics.

\textcolor{forestgreen}{\emph{\textsf{Methods}.}}---We address the electronic structures of TTG devices  
using a continuum model and a plane-wave expansion for the 
bilayer's four-component envelope function spinors.  When only one layer is twisted, whether in the middle or on the outside,  
the low-energy effective Hamiltonian is periodic and 
can be solved using Bloch's theorem. 
(In general the trilayer structure is incommensurate\cite{Christophe2019,Shi2019}.)
We choose the first Brillouin zone (BZ) shown in Fig. \ref{fig:1}(b). 
For two monolayers with a small twist angle $\theta$, the  moir{\'e} BZ  can be constructed from two monolayers' BZs. 
The monolayer BZs' neighboring corners, $K_1$ and $K_2$, where the low-energy Dirac cones of 
the isolated layers are located, form two vertices of the moir{\'e} BZ. 
The  length of the reciprocal lattice vector of the moir{\'e} pattern, $b_M$, is therefore related to that of graphene, $b_g$, 
by $b_M=2b_g\text{sin}(\theta/2)\approx b_g\theta$.   When we assume the first graphene layer is on top of the 
second layer in Fig. \ref{fig:1}(b) and add a third layer under the second layer, the third layer's BZ coincides with the 
first (second) layer's for a middle (top) layer twist and the moir{\'e} BZ is still given by Fig. \ref{fig:1}(b).

Adding tunneling between neighboring layers to the isolated layer's Dirac Hamiltonians
leads to the following effective model of TTG ( projected to the $+K$ valley):
\begin{equation}
\mathcal{H}_{\rm TTG}=\left( \begin{array}{ccc}
h_{1}(\bm k) & T_{12} & 0   \\
T_{12}^{\dagger} & h_{2}(\bm k) & T_{23}  \\
0 & T_{23}^{\dagger} & h_{3}(\bm k) \\
\end{array} \right)
\end{equation}
Here, $h_l(\bm k) (l=1,2,3)$ is the
spin-independent Dirac Hamiltonian  of the $l$th layer accounting for its orientation:
 $h_l(\bm k)=\text{diag}(e^{i\theta_l},1)(\hbar v_{\rm F}\bm k \cdot  \bm \sigma) \text{diag}(e^{-i\theta_l},1)$, 
 where $v_{\rm F}=10^6$ m/s, $\bm k$ and $\bm \sigma$ are respectively the Fermi velocity of pristine graphene,  
 a two-dimensional momentum, and the sublattice-pseudospin Pauli matrices.  
The angles, $\theta_1=-\theta/2$, $\theta_2=\theta/2$ and $\theta_3=\mp\theta/2$ for middle-layer and top-layer twists. 
The  $T_{ll'}$ interlayer tunneling terms are functions
of spatial position, $\bm r$, and have the periodicity of the moir{\'e} pattern.
In the middle-layer twist case, $T_{12}(\bm r)=w\sum^3_{m=1} \text{exp}(-i\bm {q}_m\cdot \bm r)T_m$ 
and $T_{23}(\bm r)=w\sum^3_{m=1} \text{exp}(i\bm {q}_m\cdot \bm r)T_m$.  
In the top-layer twist case, $T_{23}=w(T_1+T_2+T_3)$ is independent of position. 
Here $w$ is the tunneling energy which we set to 124.5 meV
to match infrared spectroscopy measurements \citep{Kuzmenko:2009fj}   
and the $\bm q_m (m=1,2,3)$ are the inter-layer tunneling momentum boosts,
$\bm q_{1,2}=b_M(\pm 1/2, 1/2\sqrt{3})$ and $\bm q_3=b_M(-1/\sqrt{3},0)$ \citep{Bistritzer:2011ho}. 
The matrices $T_m$ account for the sublattice dependence of tunneling
and depend on bilayer stacking: 
\begin{equation}
\begin{gathered}
T^{\rm AB}_{m}=T^{\rm BA \dagger}_{m}=\left( \begin{array}{ccc}
e^{i \frac{2m\pi}{3}} & 1  \\
 e^{-i \frac{2m\pi}{3}}& e^{i  \frac{2m\pi}{3}} 
\end{array} \right), \\
T^{\rm AA}_{m}=\left( \begin{array}{ccc}
1 & e^{-i \frac{2m\pi}{3}}  \\
e^{i \frac{2m\pi}{3}} & 1
\end{array} \right)
\end{gathered}
\end{equation}
where AB, AA and BA refer to the stacking arrangements illustrated in Fig.~\ref{fig:1}.
By using the appropriate matrices for tunneling between adjacent layers,
we can describe ABA, ABC and AAA  TTG.  

\begin{figure}[h!]
\setlength{\belowcaptionskip}{-0.2 cm}
\centering
\includegraphics[width=8.5cm]{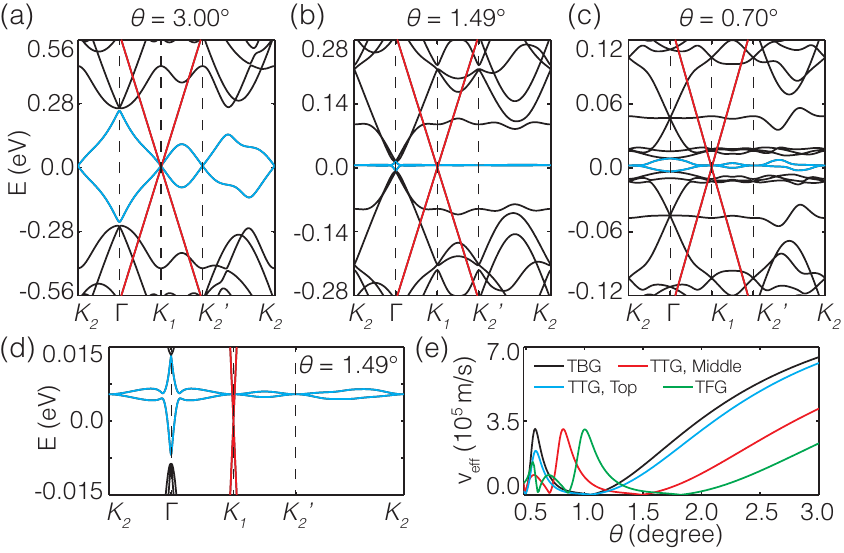}   
\caption{Band structures of  ABA-stacked TTG with a middle-layer twist, at twist angles
 (a) 3.00\textdegree, (b) 1.49\textdegree$\ $ and (c) 0.70\textdegree. (d) zooms in on low-energy bands in (b). 
 The blue and red lines are used to distinguish the odd-parity high-velocity bands from the 
 even-parity low-velocity bands. 
(e) Magnitude of the Dirac point velocity {\em vs.} twist angle. 
The black, red, blue and green lines distinguish TBG, ABA-stacked TTGs with a middle-layer twist,
ABA-stacked TTB with a top-layer twist, and twisted five-layer graphene (TFG), respectively. 
The latter structure also has bands with velocities equal to those of TBG, which are not shown.}
\label{fig:2}
\end{figure}

\textcolor{forestgreen}{\emph{\textsf{Results}.}}--- We first discuss the band structures of TTG with a middle-layer twist relative to ABA stacking, which is illustrated in Fig. \ref{fig:2}.  The top-layer twist case and some other stacking arrangements 
are discussed later for comparison. In Fig. \ref{fig:2}, three representative twist angles are chosen to illustrate
how the moir\'e bands evolve with twist.  
At a large twist angle (3.0\textdegree), we see two distinct gapless Dirac cones with 
different Fermi velocities at the band crossing momentum point, $K_1$.    
The Dirac cone with the larger Fermi velocity maintains linear  dispersion
in the illustrated 1.0 eV energy range.  One can check that its Fermi velocity is 
equal to that of an isolated graphene layer, $10^{6}$ m/s. 
In contrast, the other Dirac cone is linear only over several tens meV [See Fig. \ref{fig:2} (a)] and its dispersion is 
evidently altered by moir\'e pattern. The avoided crossing at the midpoint of $K_{1,2}$
produces a characteristic van Hove singularity \citep{Bistritzer:2011ho, Li2010, Kim:2017gc}. 
As the angle decreases,  the high-velocity Dirac cone is still undisturbed while the 
small-velocity Dirac cones at $K_1$ and $K_2$ become shallower, reaching 
velocities smaller than 200 m/s at 1.49$^\circ$. 
The  low-energy moir\'e bands becomes very flat near this magic angle,
reaching a minimum bandwidth of $\sim 20$ meV due mostly to dispersion near $\Gamma$.
Fig. \ref{fig:2} (b,d)]. At smaller angles, the renormalized velocity varies continuously, 
with more emergent magic angles [Red line in Fig. \ref{fig:2} (e)]. 
It is seen that a second magic angle appears at  0.70\textdegree, with the vanishing velocity 
and nearly flat bands  [Fig. \ref{fig:2} (c)]. 

The moir\'e band structures of TTG devices have two main 
characteristics that distinguish them from bilayers: (i) The magic angles are larger by a factor of about 1.4.  
For TBG, the first two magic angle are 1.05\textdegree$\ $and 0.50\textdegree  \citep{Bistritzer:2011ho}. 
[Fig. \ref{fig:2} (e)].  
Larger angles will make it easier to locate the magic angles and study their exotic electronic 
properties \citep{Cao:2018kn, Cao:2018ff, Wu:2018wc, Song:2018ul}. 
(ii) In stacks with mirror symmetry there is, in addition to flat bands,
one band with an undisturbed single-layer Dirac cone at 
all twist angles.  The simultaneous presence of both large and small velocity bands in a moir\'e 
superlattice opens up new opportunities for quantum simulation because of its similarity to 
circumstances that often arise in atomic-scale crystals, for example in elemental transition metals,
in which extended $s$ and localized $d$ orbitals are present in the same energy range.  

\begin{figure}[h!]
\setlength{\belowcaptionskip}{-0.2 cm}
\centering
\includegraphics[width=8.2cm]{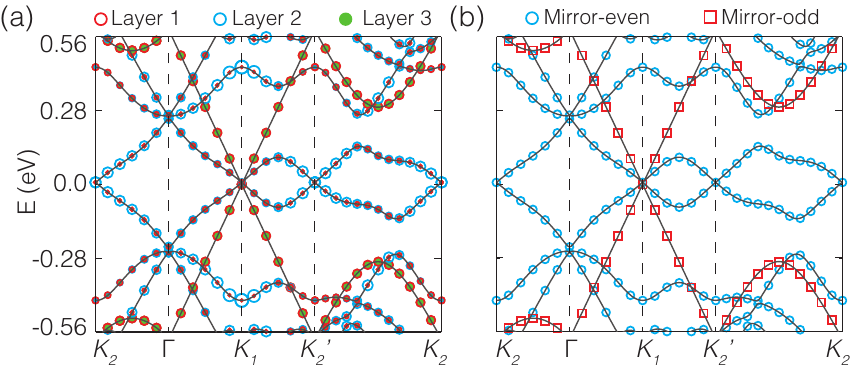}
\caption{ Layer and mirror symmetry projections of electronic states for middle-layer twist ABA TTG: (a) band structure, with the 
layer-projected weights on the first (red), second (blue) and third (green) layer indicated by the size of circles on each state. 
(b) band structures with parity projected weights indicated by blue circles (even) and red squares (odd). The twist angle $\theta= 3.0$\textdegree.}
\label{fig:3}
\end{figure}

To better understand the band structures of ABA middle-layer twist TTG devices, 
we plot the layer projection weights of band eigenstates in Fig. \ref{fig:3} (a).
It is seen that  the high-velocity Dirac cone states are on the outmost layers and have no middle layer component.
In contrast, all weight is present in all three layers for the flatter bands, implying entanglement between the middle 
and outside layers.  The outmost layer weights are always identical.  
These features can be understood in terms of the mirror symmetry of ABA stacking with respect to the middle layer.
At any twist angle, electronic states are all  eigenstates of the mirror operator. 
Fig. \ref{fig:3} (b) shows the mirror symmetry for each electronic state. 
The high-velocity Dirac cones are mirror-antisymmetric, that is, the wavefunction is odd when swapping two outmost layers.  
In the antisymmetric state, the coupling from the two outside to the middle layer interferes destructively. 
On the other hand,  the flatter bands are symmetric and their wavefuctions are unchanged under the the mirror operation.

It is useful to examine the lowest-order truncation\citep{Bistritzer:2011ho}
of the plane-wave expansion\citep{Bistritzer:2011ho},
which  allows some analytic progress by including only one Dirac cone on the outmost layers
coupled to the three nearest Dirac states of the middle layer: 
\begin{equation}
\mathcal{H}_{\rm eff}=\left( \begin{array}{ccccc}
h(\bm k) & T^{\rm AB}_{1} &  T^{\rm AB}_{2} &  T^{\rm AB}_{3} &  0    \\
T^{\rm AB\dagger}_{1}             & h (\bm k+\bm q_1)  & 0 &0 &   T^{\rm BA}_{1}\\
T^{\rm AB\dagger}_{2}             & 0     & h (\bm k+\bm q_2)   &0 & T^{\rm BA}_{2} \\
T^{\rm AB\dagger}_{3}             & 0  &0    & h(\bm k+\bm q_3)  & T^{\rm BA}_{3} \\
0             & T^{\rm BA\dagger}_{1}   &T^{\rm BA\dagger}_{2}    & T^{\rm BA\dagger}_{3}  &   h(\bm k) 
\end{array} \right)
\label{eff}
\end{equation}
Here, $h(\bm p)=\hbar v_{\text F}\bm p \cdot  \bm \sigma$ ignores the small twist.
The corresponding wavefunction is defined as $\psi=(\alpha, \beta_1,\beta_2,\beta_3, \gamma)$,
where $\alpha$, $\beta_m$ and $\gamma$ are sublattice spinors on the three layers. 
Treating the coupling to the middle layer states, which have finite energy at $\bm k=0$,  perturbatively and
using the identity $T_m h^{-1}(\bm q_m) T_m^{\dagger}=0$ \citep{Bistritzer:2011ho}, 
we find four zero energy states at $\bm k = 0$ with $\beta_m=-h^{-1}(\bm q_m) T_m^{\dagger}(\alpha+\gamma) $.
The dispersion at small $\bm k$ can then be found by diagonalizing 
$\Delta\mathcal{H}_{\rm eff}=\mathcal{H}_{\rm eff}-\mathcal{H}_{\rm eff} |_{\bm k=0}=h(\bm k) \otimes \bm I_5$ in
the $\bm k=0$ zero-energy subspace.  Reflecting the problem's mirror symmetries the 
low-energy Hamiltonian is diagonalized by $\alpha=\pm\gamma$, corresponding to mirror-symmetric and antisymmetric states.
For $\alpha=\gamma$, the low-energy dispersion, 
 \begin{equation}
\frac{\langle \psi |\Delta\mathcal{H}_{\rm eff}|\psi \rangle}{\langle \psi |\psi \rangle}=\frac{2-12\eta^2}{2+24\eta^2} \; \alpha^{\dagger} \hbar v_{\rm F}\bm k \cdot  \bm \sigma\alpha
\end{equation}
The effective Fermi velocity, $v_{\rm {eff}}$, after the renormalization is therefore 
$(2-12\eta^2)/(2+24\eta^2){v_{\rm F}}$, which depends on only one parameter $\eta=\sqrt{3}w/(\hbar v_{\text F} b_M)\varpropto\theta^{-1}$. The magic angle condition is therefore $\eta=\sqrt{1/18}$,; at this value of $\eta$ the 
expectation values of the velocity operators cancel between the outmost layers and the middle layer.
The  magic angle of TTG is  $\sqrt{2}$ time larger than that of TBG \citep{Bistritzer:2011ho}. 
The simple model captures the increase of the magic angle of TTG compared to TBG. 
The small-velocity bands act like an effective TBG system 
because the outmost layers act as a single layer.  The 
hopping energy becomes larger by a factor of $\sqrt{2}$ due to constructive interference.  
The enhanced $w$ is due to presence of more tunneling processes between the middle layer and  in-phase 
outmost layers.  For $\alpha=-\gamma$,  $\beta_m=0$ and $v_{\rm eff}=v_{\rm F}$.
Because the middle layer is absent in the mirror odd sector, 
only the outmost layers contribute to the undisturbed Dirac cone. 
The flat bands with larger magic angles and the high-velocity Dirac cone therefore 
respectively arise from the mirror-even and mirror-odd electronic sectors.  
The simple model agrees with the band projections in Fig.~\ref{fig:3} and explains the main characteristics 
in the electronic structures of middle-layer twist ABA TTG.  

\begin{figure}[h!]
\setlength{\belowcaptionskip}{-0.2 cm}
\centering
\includegraphics[width=8.5cm]{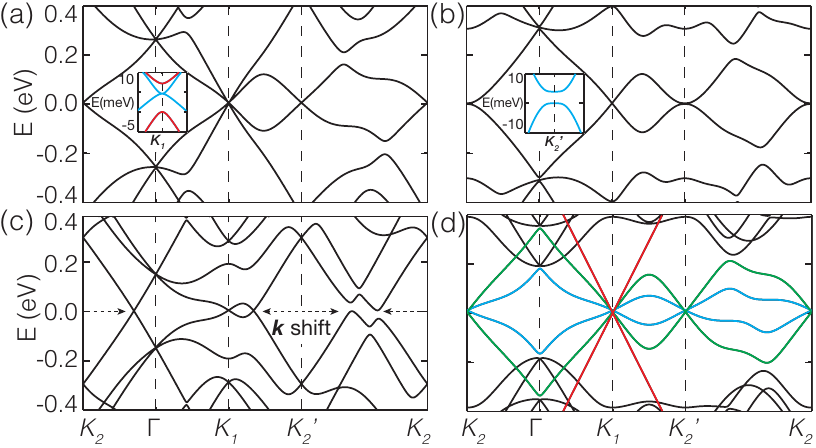}
\caption{Moir\'e band structures at twist $\theta=3.0$\textdegree \ for different twisted graphene systems: (a) ABC-stacked TTG with a middle layer twist (b) ABA-stacked TTG with a top layer 
twist and (c) AAA-stacked TTG with a top-layer twist. The insets in (a) and (b) zoom in on low-energy bands at $K_1$ and $K_2'$, respectively. 
(d) ABA-stacked TFG with colored lines distinguishing the low-energy bands.}
\label{fig:4}
\end{figure}

For TTG with a middle-layer twist, we also consider other starting stacking arrangements. 
Fig. \ref{fig:4} (a) shows the band structure that emerges from ABC stacking for middle layer twists.  In this case 
mirror symmetry is absent.  There are two Dirac cones 
with different low-energy behaviors at $K_1$, as in the ABA case.
One Dirac cone (blue bands in inset) has a reduced, but still sizable, Fermi velocity which 
reaches a minimum of $7.5\times 10^4$ m/s at 1.52\textdegree.
The other Dirac cone (red) is,  in contrast to the ABA-stacked TTG case, gapped at $K_1$.

We have also studied top-layer twist cases, which never have mirror symmetry. 
Figs. \ref{fig:4} (b) and (c) respectively  show the moir\'e  band structures for 
ABA and AAA stacking with a top layer twist.  
For  ABA stacking with a top layer twist, there are a Dirac cone at $K_1$ and a parabolic band at $K_2$ in the low-energy range, 
which at large twist angles have layer projections mainly on to the top layer and the bottom bilayer respectively.
This is expected given that $K_1$ and $K_2$ are respectively BZ corners of top layer and bottom bilayer. 
The parabolic bands are gapped [See the inset in Fig. \ref{fig:4}(b)], unlike the bands of an isolated BA-stacked bilayer \citep{Patrones:2006}, because the top layer adds a vertical potential to the bilayer \citep{Castro:2007}. 
The bandgap opening may provider opportunities for valley-contrasting optoelectronics in twisted graphene systems \citep{Xiao:2007,Yao:2008, Zhu:2018}. Moreover, the low-energy bands become flat at magic angles   
equal to  ones of TBG [Fig. \ref{fig:2}(e)], agreeing with previous results \citep{SuarezMorell:2013eta} . 
For  AAA-stacked TTG with a twist in the top layer,  the AA-stacked bottom bilayer mainly contributes to low-energy bands that are centered at $K_2$ but have a sizable momentum shift from $K_2$, similar to that of an  isolated AA-stacked bilayer    \citep{Andres:2008}. The momentum differences between Dirac cone at $K_1$ and low-energy bands around $K_2$ are no longer equal to 
$\bm q_m$, and  interlayer tunnelings between these states are suppressed.  
Flat bands and vanishing velocities  are therefore not found for this stacking.   

\textcolor{forestgreen}{\emph{Discussion}}--- When one layer is twisted relative to another, TBG
forms a moir\'e superlattice in which regions with local AA, AB and BA stacking form three offset triangular lattices. 
The stacking (for example AA, AB or BA) prior to the twist changes the twist axis but otherwise lead to the same moir{\'e} pattern.
Indeed, the electronic properties of TBG are independent of the original stacking \citep{Bistritzer:2011ho,Wu:2018wc}. 
Any graphene multilayer with more than two layers also forms a periodic moir\'e superlattice when one layer, or one subset 
of layers, is rotated relative to the other layers.  Each is characterized by a spatially varying local multilayer stacking arrangement 
which varies periodically in space. Focusing on the trilayer case, we have demonstrated that a wide variety of 
different behaviors result from small twists, depending on the original stacking order and on which layer is twisted.  
In the multilayer case, the moir{\'e} supperlattice electronic properties vary widely even for 
the same twist angle. It is because different original  stacking orders and twist layers result in different moir{\'e} patterns that determine  electronic  properties. 
That is, the moir\'e superlattice of the TTG, as well as twisted few-layer graphene, has no longer one pattern, distinct from that of TBG.  In particular, moir{\'e} patterns with and without mirror symmetry demonstrate starkly different electronic properties.
Moreover, for a middle-layer twist, local ABA  and AAA stackings coexist in the moir{\'e} pattern with mirror symmetry.
 Therefore,  the TTG formed by a rotation relative to AAA stacking  has the same geometrical and electronic structures 
 with the ABA-stacked TTG, as shown in Fig. \ref{fig:2}. The same applies to ABC and BAA stackings that both appear in the moir{\'e} pattern without mirror symmetry.
 Similarly, for a top-layer twist,  BBA- and BAA-stacked TTGs have the same electronic structures with ABA- and AAA-stacked ones, respectively.

Compared with TBG, the magic angles in a mirror-protected TTG  increase by a factor of $\sqrt{2}$. 
This increase can be enhanced by increasing the thickness of the stacks.
We computed the electronic structure of twisted five-layer graphene (TFG) structures 
starting from ABABA Bernal stacking and rotating even-numbered layers relative to odd-numbered layers.
In this case the middle layer is a mirror plane.  As shown in Fig. \ref{fig:4} (d), there are  two low-velocity Dirac cones at $K_1$ 
(blue and green bands) with velocity renormalization and one undisturbed Dirac cone (red) with
the same velocity as an isolated graphene layer. The two modified Dirac cones yield two groups of magic angles,
one larger by a factor of $\sqrt{3}$ yielding a largest magic angle of 1.82\textdegree, as illustrated in  
Fig. \ref{fig:2} (e), and another set of magic angles equal to those of TBG. 
The possibility of larger magic angles and the coexistence between flat bands and undisturbed Dirac cones 
occur in mirror-protected TFG, just as in like TTG.  
These electronic properties are well explained by the truncated Hamiltonians like Eq. \ref{eff} (See details in Appendix A). 
The larger magic angles arise from  the coherent enhancement of stacked TBGs in a mirror-even eigenstate, similar to TTG.  
Similarly twisted seven-layer  graphene is computed to have two magic angles larger than 1\textdegree, at  1.49\textdegree$\ $and 1.94\textdegree.  For twisted nine-layer,  the magic angles larger than 1\textdegree$\ $ are 1.24\textdegree, 1.70\textdegree$\ $and 2.00\textdegree. 

The possibility of more and larger magic angles in twisted few-layer graphenes provides a powerful 
motivation for fabricating and characterizing these multilayer devices to explore their flat-band-related physics.
The coexistence of small and large velocity bands in TTG with mirror symmetry could have important implications 
for superconductivity when effective interactions between quasiparticles are attractive, and for 
itinerant electron magnetism when effective interactions between quasiparticles are repulsive.  
In the superconducting case the undisturbed Dirac cone with high velocity could lead to enhanced superfluid stiffness and increases
in the the Berezinskii-Kosterlitz-Thouless transition temperature.  In the magnetic case, the high velocity band could 
stiffen magnons and again increase ordering temperatures.  
This exploratory theoretical work makes it clear that the study of many-body interaction effects in
multilayer graphene moir\'e superlattices is still at an early stage.  

\bigskip









\begin{acknowledgments}
\noindent\textbf{Acknowledgments} X. L. is grateful to Bangguo Xiong for valuable discussions.
A. M. is supported by the Welch Foundation under Grant No. TBF1473 and the Department of Energy under Grant No. DE-FG02-ER45118. X. L. is supported by the start-up funds from Nanjing Normal University and  F. W. is supported by Laboratory for Physical Sciences.
\end{acknowledgments}

\section{Appendix A: The simplest model of twisted five-layer graphene}
To  investigate the origin of low-energy bands of  twisted five-layer graphene in Fig.  \ref{fig:4} (d), we extend the effective model in Eq. \ref{eff} and make the submatrix for the third-fifth layers  equal to that for the first-third layers. We define the wavefuction as $\psi=(\alpha, \beta_1,\beta_2,\beta_3, \gamma, \delta_1,\delta_2, \delta_3, \epsilon)$, where $\alpha$, $\beta_m$, $\gamma$, $\delta_m$ and $\epsilon$ are spinors acting on the sublattices of  the first to fifth layers, respectively.  According to the relation $T_m h^{-1}(\bm q_m) T_m^{\dagger}=0$,  three  wavefunction's  expressions for the zero-energy eigenvalues at $\bm k=0$ are obtained with $h(0)\alpha=h(0)\beta=h(0)\epsilon=0$. They correspond to two low-velocity Dirac cones with the velocity renormalization and one undisturbed Dirac cone with isolated graphene's Fermi velocity, as followings,

(i) A mirror-even wavefunction has $\alpha=\gamma/2=\epsilon$ and $\beta_m=\delta_m=-3h^{-1}(\bm q_m) T_m^{\dagger}\alpha$. The low-energy dispersion of the Dirac bands is then computed by the expectation value of  $\Delta\mathcal{H}'_{\rm eff}=h(\bm k) \otimes \bm I_9$, that is,  
\begin{equation}
\frac{\langle \psi |\Delta\mathcal{H}'_{\rm eff}|\psi \rangle}{\langle \psi |\psi \rangle}=\frac{6-54\eta^2}{6+108\eta^2}\alpha^{\dagger} \hbar v_{\rm F}\bm k \cdot  \bm \sigma\alpha
\end{equation}
Therefore, the renormalized Fermi velocity 
$v_{\rm {eff}}=(6-54\eta^2)/(6+108\eta^2){v_{\rm F}}\approx (1-27\eta^2){v_{\rm F}}$. The magic angle appears at $\eta=\sqrt{1/27}$ where $v_{\rm {eff}}=0$. The angle increases by a factor of $\sqrt{3}$ compared to TBG, due to coherent enhancement of stacked TBGs under the action of mirror symmetry.  

(ii) A mirror-odd state corresponds to $\alpha=-\epsilon$, $\gamma=0$ and $\beta_m=-\delta_m=-h^{-1}(\bm q_m) T_m^{\dagger}\alpha$. The vanishing contribution from the middle layer  isolates the top TBG (i.e. the first and second layers) and the bottom TBG (the fourth and fifth layers).  In a similar way as the above wavefunction, the calculated Fermi velocity, $v_{\rm {eff}}=(1-3\eta^2)/(1+6\eta^2){v_{\rm F}}\approx (1-9\eta^2){v_{\rm F}}$,  according to the expectation value of  $\Delta\mathcal{H}'_{\rm eff}$. The corresponding magic angle appears at $\eta=1/3$, which is  equal to that of TBG.  

(iii) The other mirror-even state gives $\alpha=-\gamma=\epsilon$ and $\beta_m=\delta_m=0$. Similar to the above solution, the absence of the second and fourth layers' components separate three odd-numbered monolayers, leading to  isolated graphene's Dirac cone.


\end{document}